\documentclass[letterpaper, 10 pt, conference]{ieeeconf}  

\IEEEoverridecommandlockouts                              
\usepackage{amsmath}
\usepackage{amssymb}
\usepackage{cite}
\usepackage{stfloats}
\usepackage{amsthm}
\usepackage{indentfirst}
\usepackage{url}
\usepackage{array}
\usepackage[utf8]{inputenc}
\usepackage[english]{babel} 
\usepackage{graphicx}
\usepackage{epstopdf}
\graphicspath{ {images/} }
\usepackage{float}
\usepackage{wasysym}
\usepackage{multirow}
\usepackage{mathrsfs}
\usepackage{enumerate}
\usepackage{color}
\usepackage{slashbox}
\usepackage[ruled,vlined]{algorithm2e}
\usepackage{subcaption}
\usepackage{pdfpages}

\theoremstyle{definition}
\newtheorem{remark}{Remark}

\newtheorem{theorem}{Theorem}

\title{\LARGE \bf
A Stackelberg Security Investment Game for Voltage Stability \\ of Power Systems
}

\author{Lu An, Aranya Chakrabortty, and Alexandra Duel-Hallen
\thanks{The authors are with the Department of Electrical and Computer Engineering, North Carolina State University, Raleigh, NC 27695. (e-mail: {\tt\small lan4@ncsu.edu}, {\tt\small achakra2@ncsu.edu}, {\tt\small sasha@ncsu.edu}) This paper has been accepted by IEEE CDC 2020.}
}

\begin{document}

\maketitle
\thispagestyle{empty}
\pagestyle{empty}

\begin{abstract}
We formulate a Stackelberg game between an attacker and a defender of a power system. The attacker attempts to alter the load setpoints of the power system covertly and intelligently, so that the voltage stability margin of the grid is reduced, driving the entire system towards a voltage collapse. The defender, or the system operator, aims to compensate for this reduction by retuning the reactive power injection to the grid by switching on control devices, such as a bank of shunt capacitors. A modified 
Backward Induction method is proposed to find a cost-based Stackelberg equilibrium (CBSE) of the game, which saves the players' costs while providing the optimal allocation of both players’ investment resources under budget and covertness constraints. We analyze the proposed game extensively for the IEEE 9-bus power system model and present an example of its performance for the IEEE 39-bus power system model. It is demonstrated that the defender is able to maintain system stability unless its security budget is much lower than the attacker's budget.
\end{abstract}
\begin{keywords} Stackelberg game, voltage stability, load attacks, security investment, power systems\end{keywords}

\section{Introduction}
Over the past decade, significant research has been done on
cyber-security of power systems \cite{anu} with applications in state
estimation \cite{peng}, volt/VAr control \cite{dan3}, automatic
generation control \cite{ashok}, etc. In this paper we address an equally important and yet less-visited cyber-attack scenario in power systems - namely, covert attacks on loads causing degradation of voltage stability. Unlike other papers, such as \cite{rad}, that report detection and control problems for load attacks, our goal is to formulate an investment strategy that power system operators can adopt to secure the grid when an attacker tries to drive it to voltage collapse by manipulating a chosen set of load setpoints. This manipulation can be done in a covert way for each individual load, so that the user does not feel any difference in consumption, but when hundreds of such loads are tweaked simultaneously, the cumulative effect can still result in severe degradation of voltage stability \cite{simpson2016voltage}.

We use a Stackelberg game (SG) \cite{Osborne1994} to formulate this security investment, considering the defender as the leader and the attacker as the follower \cite{Basar2019}. Game theory has been a common tool for analyzing security problems in cyber-physical systems \cite{7011006,Basar2019}. Cooperative and non-cooperative games have also been proposed for non-attack scenarios such as load balancing and voltage stability in \cite{Avraam2018VoltageCS, 8107250}. To the best of our knowledge, no research has been done to explore how game theory pertains to security investment for load attacks. Moreover, most game-theoretic security investment research employs dynamic games \cite{Basar2019}, including stochastic games and games that utilize learning, where the players repeatedly update their investment strategies in response to the opponents’ actions. However, these games are not practical when long-term, fixed security investment is desired.

The SG for our problem is set up as follows. The attacker plans to hack covertly into a set of loads and modify their setpoints to increase the system-wide voltage instability index \cite{simpson2016voltage}. The defender switches on control devices to compensate the reactive power balance in the grid proactively, so that the instability index remains close to its safe value if an attack occurs in the future. Both players are subject to budget constraints. Additionally, the attacker aims to remain covert. We modify the standard backward induction \cite{Osborne1994} for SG to choose a Stackelberg equilibrium (SE) \cite{amir1999stackelberg} that reduces the players' costs while retaining their payoffs. The resulting cost-based Stackelberg equilibrium (CBSE) provides guidelines to the system operator for fixed, long-term grid protection against voltage stability attacks. We validate our results using the IEEE 9-bus and 39-bus power system models and demonstrate that voltage stability can be maintained unless the defender's security resources are much more limited than the attacker's budget.

\section{Power System Model} \label{sec:sysmod}
We first recall the definition of voltage instability index from \cite{simpson2016voltage}, which will be used as the central metric for evaluation of our game.  Consider a power system with $M \geq 1$ generators, and  $K \geq 1$ loads, where the load buses are indexed as the first $K$ buses, followed by $M$ generator buses.  Let the steady-state voltage magnitudes at the load buses be stacked as ${{\mathbf{V}}_L} = [V_1,\cdots,V_K] \in {\mathbb R^K}$, and at the generator buses as ${{\mathbf{V}}_G} = [V_{K+1},\cdots,V_{K+M}] \in {\mathbb R^M}$. Let the admittance matrix of the network be denoted as ${\mathbf{Y}}={\mathbf{G}}+j{\mathbf{B}}$, where ${\mathbf{B}}$ is referred to as the susceptance matrix. We partition the susceptance matrix ${\mathbf{B}} \in \mathbb R^{(K +M) \times (K +M)}$ into four block matrices as:
\begin{equation}
\label{eq:B}
{\bf{B}} = \left( {\begin{array}{*{20}{c}}
{{{\bf{B}}_{LL}}}&{{{\bf{B}}_{LG}}}\\
{{{\bf{B}}_{GL}}}&{{{\bf{B}}_{GG}}}
\end{array}} \right),
\end{equation}
where ${{\bf{B}}_{LL}}$ contains the interconnections among loads, and ${{\bf{B}}_{LG}}={{\bf{B}}_{GL}}^{T}$ represents the interconnections between loads and generators. Following the derivations in \cite{simpson2016voltage}, one can then define the open-circuit load voltage vector as:
\begin{equation}
\label{eq:VLo}
{\bf{V}}_L^* =  - {\bf{B}}_{LL}^{ - 1}{{\bf{B}}_{LG}}{{\bf{V}}_G},
\end{equation}
and, subsequently, the symmetric stiffness matrix as:
\begin{equation}
\label{eq:Q_cirt}
{{\bf{Q}}_{cirt}} \triangleq \frac{1}{4}{\rm{diag}}({\bf{V}}_L^*) \cdot {{\bf{B}}_{LL}} \cdot {\rm{diag}}({\bf{V}}_L^*),
\end{equation}
where $\rm{diag}(\cdot)$ denotes the diagonal matrix.

Using (\ref{eq:VLo}) and (\ref{eq:Q_cirt}), the \textit{voltage instability index} of the system can be defined as:
\begin{equation}
\label{eq:delta}
\Delta  = ||{\mathbf{Q}}_{cirt}^{ - 1}{{\mathbf{Q}}_L}|{|_\infty },
\end{equation}
where ${{\mathbf{Q}}_L} = [Q_1,\cdots,Q_K] \in \mathbb R^K$ is a $K$-dimensional real vector that represents the \textit{reactive power setpoints} at the load buses. Here, $||\cdot||_\infty$ refers to the $\ell_{\infty}$-norm, which picks the absolute value of the element with the largest magnitude in a vector. The $k^{\text{th}}$ entry of the matrix-vector product ${\mathbf{Q}}_{cirt}^{ - 1}{{\mathbf{Q}}_L}$ captures the stability stress on load $k$, with $||\cdot||_\infty$ identifying the maximally stressed node. According to Theorem 1 in \cite{simpson2016voltage}, the power flow equation will have a unique, stable solution if $\Delta<1$. Equivalently, $\Delta\geq1$ indicates that at least one load bus in the system is overly stressed and can be responsible for a voltage collapse. We refer to $1-\Delta$ as the \textit{voltage stability margin} \cite{van2007voltage}. The larger the value of $\Delta$, the narrower the stability margin is and the closer the power system is to a voltage collapse. Denote the \textit{nominal voltage stability index} ${\Delta ^0}$ as the value of $\Delta$ computed from (\ref{eq:delta}) using the \textit{nominal reactive power setpoints} ${{\mathbf{Q}}_L^0}$ (over a certain period of time assuming that the setpoints are constant over this period).

According to Proposition 3 in supplementary note 6 of \cite{simpson2016voltage}, $\mathbf{Q}_{cirt}^{-1}$ has negative elements and ${\mathbf{Q}_L}$ has positive elements. Thus, if some elements of ${\mathbf{Q}_L}$ increase, the $\ell_{\infty}$-norm in (4) also increases. Therefore, the voltage instability index $\Delta$ in (4) increases as the reactive power demands of the loads grow. An attacker can increase the reactive power demands at appropriately chosen load buses by adding an incremental vector ${{\mathbf{q}}_a} = [q_a^1,\cdots,q_a^K] \in \mathbb R^K$ to ${{\mathbf{Q}}_L^0}$ and thus easily narrow down the voltage stability margin. Since only the reactive power setpoints are tampered with, and not the active power setpoints, the user may not feel any difference in her consumption pattern, which makes this type of attack unobservable to a large extent. The attacker can further make this attack covert by designing the entries of ${{\mathbf{q}}_a}$ small enough that they maintain the load bus voltages to be within their usual allowable range of 0.9 per unit (pu) to 1.1 pu while still pushing $\Delta$ towards 1. To prepare for possible future attacks, the operator, or the defender, can switch on voltage control devices, such as shunt capacitors and power electronic converters, to compensate for the potential increase in consumption in advance. These control devices may or may not be located at the load bus. If they are not, their equivalent contribution of reactive power at the $K$ load buses can be obtained by simple network reduction. Let this equivalent $K$-dimensional reactive power compensation vector be denoted as ${{\mathbf{q}}_d}=[q_d^1,\cdots,q_d^K] \in \mathbb R^K$. When an attack happens, the overall reactive power balance becomes ${{\bf{Q}}_L^{'}}={{{\bf{Q}}_L^0} + {\bf{q}}_a - {{\bf{q}}_d}}$. The goal of the defender is to compensate for the attacker's actions and to avoid the voltage collapse by maintaining the post-attack $\Delta$ as close as possible to the nominal $\Delta^0$. We assume that the players have full knowledge of the system model and each other’s parameters. Thus, this investigation characterizes ideal game performance. We plan to extend it to uncertain scenarios in future work.

\section{The Cost-based Stackelberg Game} \label{sec:SG}
In the proposed {\it Stackelberg game (SG)}, the \textit{actions} of the attacker, $\mathbf{a} \in \mathbb R^K$, and the defender, $\mathbf{d} \in \mathbb R^K$, correspond to a finite number of discrete \textit{investment levels} into the $K$ loads and $K$ control devices, respectively. A higher value of each element $a_k$ (or $d_k$) indicates a greater chance of successful attack (or protection) of the $k^{\text{th}}$ load. Given an investment pair (${\mathbf{a}}$, ${\mathbf{d}}$), the \textit{utilities}, or \textit{payoffs}, of the attacker and the defender are termed ${U^a}({\mathbf{a}},{\mathbf{d}})$ and ${U^d}({\mathbf{a}},{\mathbf{d}})$, respectively, expressed in terms of the instability index $\Delta$. The attacker aims to maximize $\Delta$ (thus degrading the system performance) while the defender aims to reduce it. In this zero-sum game \cite{Osborne1994}, ${U^d}({\mathbf{a}},{\mathbf{d}}) = - {U^a}({\mathbf{a}},{\mathbf{d}})$. The defender is the \textit{leader}, who establishes its investment profile first. Given a defenders' strategy ${\mathbf{d}}$, the attacker \textit{follows} by choosing its action ${{\mathbf{a}}}=g({\mathbf{d}})=\mathop {\arg \max }\limits_{{\mathbf{a}}} {U^a}({{\mathbf{a}}},{{\mathbf{d}}})$, a best response to ${\mathbf{d}}$. Thus, the defender chooses a strategy ${{\mathbf{d}}^*}$ that maximizes its utility given the attacker's best responses $g({\mathbf{d}})$ to all its actions. A resulting Stackelberg equilibrium (SE) \cite{Osborne1994} (${{\mathbf{a}}^*}$, ${{\mathbf{d}}^*}$), where ${{\mathbf{a}}^*}=g({{\mathbf{d}}^*})$, optimizes the utility of each player in an SG. Finally, we modify the standard \textit{Backward Induction} (BI) method \cite{Osborne1994} for computing an SE and develop the \textit{cost-based Stackelberg game} (CBSG) that saves the players' costs without compromising their payoffs.



\subsection{Players' Actions and Cost Constraints}
The attacker's actions are denoted as ${\mathbf{a}}=[a_1,\cdots, a_k, \cdots, a_K]  \in \mathbb R^K$, where $a_k 
\in \{0,1/(L_a-1), \allowbreak 2/(L_a-1),\cdots,1\}$ is a discrete level of investment into load $k$, and $L_a$ denotes the number of attacker's investment levels. We assume each load is equipped with protective software. The value of
$a_k$ denotes the probability of successfully hacking into load $k$, which is determined by attacker's investment level, or the amount of resources allocated to hacking this load. Thus, for any attack action ${\mathbf{a}}$, there are $2^K$ possible outcomes. Define the $i^{\text{th}}$ outcome of attack at all loads by a binary $K$-tuple $\mathbf{O}^i = [o_1^i, \cdots, o_k^i, \cdots, o_K^i], \forall i=1,\cdots,2^K$, where $o_k^i=1$ if attack at node $k$ is successful and $o_k^i=0$ if it fails. Given an attacker's action vector ${\mathbf{a}}$, the probability of outcome $\mathbf{O}^i$ is given by:
\begin{equation}
\label{eq:Pr_Oi}
P_{{\mathbf{a}}}(\mathbf{O}^i) = \prod\limits_{k:{\forall o_k^i} = 1} {{a_k}} \prod\limits_{k:{\forall o_k^i} = 0} {\left( {1 - {a_k}} \right)}.
\end{equation}

In addition, we assume that if the attacker successfully hacks into load $k$, the nominal reactive power demand $Q_k$ of this load will be increased by $q_a^k$, where $Q_k$ is the $k^{\text{th}}$ element of ${{\mathbf{Q}}_L}$ in (\ref{eq:delta}). The combined incremental demand for outcome $\mathbf{O}^i$ is represented by a $1 \times K$ vector given by:
\begin{equation}
\label{eq:q_a_i}
{\mathbf{q}}_a^i = \mathbf{O}^i \odot {{\mathbf{q}}_a},
\end{equation}
where ${{\mathbf{q}}_a}=[q_a^1,\cdots,q_a^k,\cdots,q_a^K]$ and $\odot$ indicates element-wise multiplication.

Next, we define the defender's actions as $\mathbf{d}=[d_1,\cdots,d_k,\cdots,d_K] \in \mathbb R^K$, where $d_k\in\{0,1/(L_d-1),2/(L_d-1),\cdots,1\}$ denotes the defender's investment level on load $k$, or equivalently, the control device of that load, and $L_d$ is the number of defender's investment levels. Let us assume the maximum reactive power that the defender is able to compensate on load $k$ is $q_d^{k,\max}$ when the level $d_k=1$, where $q_d^{k,\max}$ is selected so that the voltage at that load bus does not exceed 1.1 pu. For the level $d_k$, the defender's compensation is $q_d^k= d_k q_d^{k,\max}$. The reactive power demand compensation for all loads is specified by the $1 \times K$ vector:
\begin{equation}
\label{eq:q_d}
{{\mathbf{q}}_d}=[q_d^1,\cdots,q_d^k,\cdots,q_d^K].
\end{equation}

Finally, we assume both players' investments are subject to the following constraints. The \textit{attacker's constraints} include:
\begin{enumerate}
\item \textbf{Cost constraint}: Assume attack on load $k$ at full effort (i.e., when $a_k=1$) has cost $\gamma_a  \in \mathbb R$. Scaling this cost by the level of effort $a_k$ and summing over all loads, we obtain the following constraint on the total cost of the attacker:
\begin{equation}
\label{eq:a_ac}
\gamma_a{||{{\mathbf{a}}}||_1} \le 1,
\end{equation}
where $||\cdot||_1$ denotes the $\ell_1$-norm, which is given by the sum of the magnitudes of all elements of the vector.
\item \textbf{Covertness constraint}: Considering that the voltage at every load bus is mandated to be within an operating range of 0.9 to 1.1 pu, the attacker must be covert in the sense that it cannot change the demand at any target bus $k$ beyond a limit $q_a^{k,\max}$ as that may violate this voltage range, leading to the attack being caught by the operator. This covertness constraint is, therefore, modeled as:
\begin{equation}
\label{eq:a_cc}
q_a^k \leq q_a^{k,\max}, \forall k.
\end{equation}
Note that $q_a^{k,\max}$ will be different for different $k$ due to physical variabilities of the loads.
\end{enumerate}

The \textit{defender} has the following \textit{constraint}:
\begin{enumerate}
\item[] \textbf{Cost of protection}: Assuming full protection (i.e. $d_k=1$) for load $k$ costs $\gamma_d  \in \mathbb R$, the defender's budget constraint is given by:
\begin{equation}
\label{eq:d_pc}
\gamma_d{||{{\mathbf{d}}}||_1} \le 1.
\end{equation}
\end{enumerate}
\begin{remark}
In (\ref{eq:a_ac}) and (\ref{eq:d_pc}), we assumed without loss of generality that the total cost of each player is bounded by 1. Thus, the scalars $\gamma_a$ and $\gamma_d$ represent scaled costs per load of the attacker and defender, respectively.
\end{remark}

\subsection{Players' Utility Functions}
Prior to the attack, the instability index $\Delta = \Delta^0$. The attacker aims to increase $\Delta$, but not exceed $\Delta=1$ since the latter results in system voltage collapse and any additional investment wastes the attacker's resources. Moreover, to save its cost, the defender invests only to compensate for the attacker's action, i.e. it aims to reduce $\Delta$ while maintaining $\Delta\geq \Delta^0$. Thus, the utilities of the players are defined in terms of the \textit{deviation} $\Delta - {\Delta ^0}$.

Given the attacker's and defender's actions ${\mathbf{a}}$ and ${\mathbf{d}}$, respectively, the reactive power demand vector for the $i^{\text{th}}$ outcome $\mathbf{O}^i$ is computed as ${{\bf{Q}}_L^{i'}}={{{\bf{Q}}_L^0} + {\bf{q}}_a^i - {{\bf{q}}_d}}$. The \textit{attacker's utility} for the $i^{\text{th}}$ outcome $\mathbf{O}^i$ is given by:
\begin{equation}
\label{eq:Ua_i}
U_i^a({\bf{d}}) = Clip\left( {{{\left\| {{\bf{Q}}_{cirt}^{ - 1}{\bf{Q}}_L^{i'}} \right\|}_\infty }}; (\Delta^0, 1) \right) - {\Delta ^0},
\end{equation}
where
\begin{equation}
\label{eq:clip}
Clip(x;({\Delta ^0},1)) = \left\{ {\begin{array}{*{20}{c}}
{{\Delta ^0}}&{x \le {\Delta ^0}}\\
x&{{\Delta ^0} < x < 1}\\
1&{x \ge 1}
\end{array}} \right..
\end{equation}

Given the strategy pair $({{\mathbf{a}}}, {{\mathbf{d}}})$ under the attacker's constraints (\ref{eq:a_ac}) and (\ref{eq:a_cc}), the attacker's utility is represented as the expectation of (\ref{eq:Ua_i}) over all outcomes:
\begin{align}
\label{eq:Ua}
{U^a}({\bf{a}},{\bf{d}}) &= {\rm E}\left( {U_i^a({\bf{d}})} \right) = \sum\limits_i^{{2^K}} {{P_{{\mathbf{a}}}(\mathbf{O}^i)U_i^a({\bf{d}})} } , \hfill  \\
 \mbox{s.t.}\;\;\;{\kern 1pt} & \gamma_a{||{{\mathbf{a}}}||_1} \le 1, \;\;
  q_a^k \leq q_a^{k,\max}, \forall k. \nonumber
\end{align}
In the proposed zero-sum game, the \textit{defender's utility} under the constraint (\ref{eq:d_pc}) is given by:
\begin{align}
\label{eq:Ud}
{U^d}({\bf{a}},{\bf{d}}) & = -{U^a({\bf{a}},{\bf{d}})},  \hfill \\
  \mbox{s.t.}\;\;\;{\kern 1pt} & {\gamma_d}||{{\mathbf{d}}}||_1 \le 1.\nonumber
\end{align}

Finally, we make the following realistic assumptions: (i) the attacker
is able to cause voltage collapse when it has unlimited resources and
the defender is inactive, and (ii) the defender is able to compensate
fully for the attacks when both players have unlimited budgets.



\subsection{Cost-based Stackelberg Equilibrium (CBSE)}
An SE is usually found using the \textit{Backward Induction} (BI) algorithm \cite{Osborne1994}. Since multiple SEs are possible in an SG, we modify the BI method to select an SE that saves both players' costs. The \textit{Cost-based Backward Induction} (CBBI) algorithm is described below:\\
\textbf{Step 1:} \textbf{(a)} For each defender's action ${\mathbf{d}}$ that satisfies (\ref{eq:d_pc}), the attacker determines the set of its best responses $\mathcal{G}({\mathbf{d}})$, where ${g}({{\mathbf{d}}}) \in \mathcal{G}({\mathbf{d}})$ if
\begin{align}
\label{eq:a_r1}
{g}({{\mathbf{d}}}) &= \mathop {\arg \max }\limits_{{\mathbf{a}}} {U^a}({{\mathbf{a}}},{{\mathbf{d}}}), \hfill  \\
 \mbox{s.t.}\;\;\;{\kern 1pt} & \gamma_a{||{{\mathbf{a}}}||_1} \le 1, \; q_a^k \leq q_a^{k,\max}, \forall k, \nonumber 
\end{align}

\textbf{(b)} For any ${\mathbf{d}}$, if there are multiple attacker's best responses in  $\mathcal{G}({\mathbf{d}})$, the attacker chooses a response with the \textit{smallest cost}:
\begin{equation}
\label{eq:a_r2}
g_o({\bf{d}}) = \mathop {\arg \min }\limits_{{g}({\bf{d}}) \in \mathcal{G}({\mathbf{d}})} ||{g}({\bf{d}})||_1.
\end{equation}

\textbf{Step 2: (a)} The defender determines the set of investment strategies $\mathcal{D}$ that maximize its payoff where ${{\mathbf{d}}^*} \in \mathcal{D}$ if:
\begin{align}
\label{eq:d_r1}
{{\mathbf{d}}^*} &= \mathop {\arg \max }\limits_{\mathbf{d}} {U^d}({g_o}({\mathbf{d}}),{\mathbf{d}}),  \hfill\\
  \mbox{s.t.}\;\;\;{\kern 1pt} & {\gamma_d}||{{\mathbf{d}}}||_1 \le 1. \nonumber
\end{align}

\textbf{(b)} If multiple solutions exist in $\mathcal{D}$, a strategy with the \textit{smallest cost} is chosen:
\begin{equation}
\label{eq:d_r2}
{{\mathbf{d}}_o^*} = \mathop {\arg \min }\limits_{{{\mathbf{d}}^*}\in \mathcal{D}} ||{{\mathbf{d}}^*}||_1.
\end{equation}
Denote
\begin{equation}
\label{eq:a_r3}
\mathbf{a}_o^* = g_o({\mathbf{d}}_o^*).
\end{equation}

The strategy pair $({\mathbf{a}_o^*},{\mathbf{d}_o^*})$ in (\ref{eq:d_r2}) and (\ref{eq:a_r3}) is a \textit{cost-based Stackelberg equilibrium (CBSE)}, and the corresponding game is termed the \textit{cost-based Stackelberg game (CBSG)}. The following Theorem summarizes several properties of SGs and of the proposed CBSG.

\theorem            \ \\
\textbf{(a)} An SE exists in a finite two-player SG.
\\
\textbf{(b)} All SEs of a zero-sum SG have the same payoffs.
\\
\textbf{(c)} In a CBSG, given $L_a$ and $L_d$, the utility of each player is non-increasing with its cost per load when the opponent's cost per load is fixed.
\\
\textbf{(d)} Given $L_a$ and $L_d$, there exist $\epsilon>0$ and $\theta>0$ such that when $\gamma_a<\epsilon$ while $\gamma_d>\theta$, the attacker's utility at CBSE $U^a({\mathbf{a}_o^*},{\mathbf{d}_o^*}) = 1-\Delta^0$ (i.e. voltage collapse occurs). Moreover, there exists an $\alpha>0$ such that when $\gamma_d<\alpha$, the attacker's utility at CBSE $U^a({\mathbf{a}_o^*},{\mathbf{d}_o^*}) = 0$ (i.e. $\Delta=\Delta^0$).
\\
\textbf{(e)} When $L_d$ (or $L_a$) is increased to a number of investment levels $L_d^{'}$ (or $L_a^{'}$) that satisfies $L_d^{'}-1=n(L_d-1)$ (or $L_a^{'}-1=n(L_a-1)$), where $n$ is a positive integer, the defender's (or attacker's) utility does not decrease if the costs and the opponent's number of investment levels $L_a$ (or $L_d$) are fixed.

\begin{proof}
Please refer to \cite{AnCDCsupp}, \cite[Appx.B]{ An2020thesis}.
\end{proof}

From Theorem 1, CBBI selects an SE with reduced costs of both players while providing the payoff of any other SE.
\begin{remark}
Instead of defining a zero-sum game with hard cost constraints, a general-sum SG, where the costs and covertness are incorporated into in the utility functions \cite{8815018}, can be investigated. Note that Theorem 1(b) does not hold for this non-zero sum game. We expect the performance trends of this game to resemble those of the proposed zero-sum SG.
\end{remark}

\section{Numerical Results}
\label{sec:cs}
\subsection{Game Analysis for the IEEE 9-bus System}
\begin{figure}[H]
  \centering
  \vspace{-0.1in}
    \includegraphics[width=0.35\textwidth]{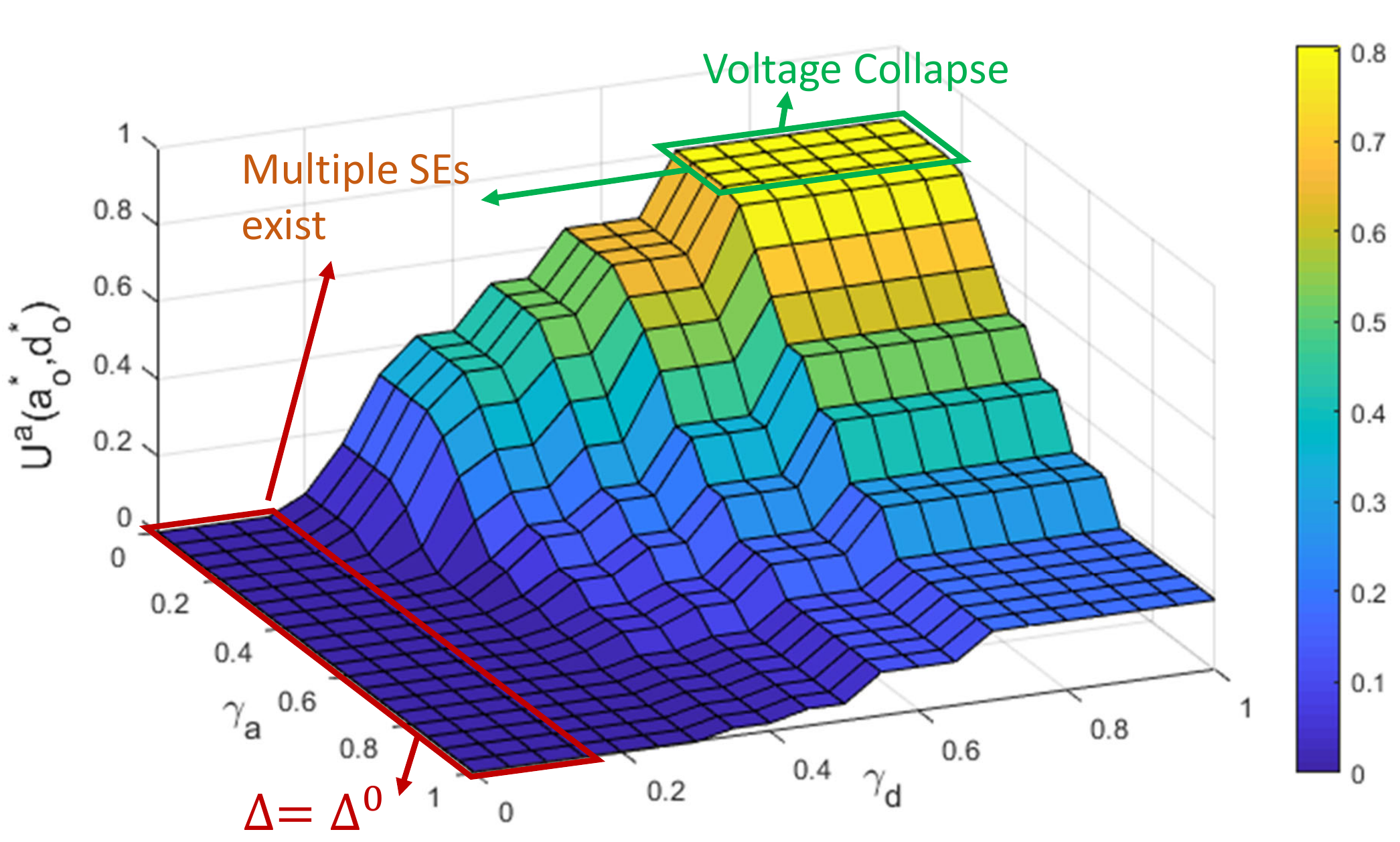}
    \caption{Attacker's utility at CBSE vs. $\gamma_a$ and $\gamma_d$ for $L_a=L_d=3$}
    \label{fig:Ua9_La3Ld3}
    \vspace{-0.2in}
\end{figure}

\begin{figure}[H]
  \centering
    \includegraphics[width=0.47\textwidth]{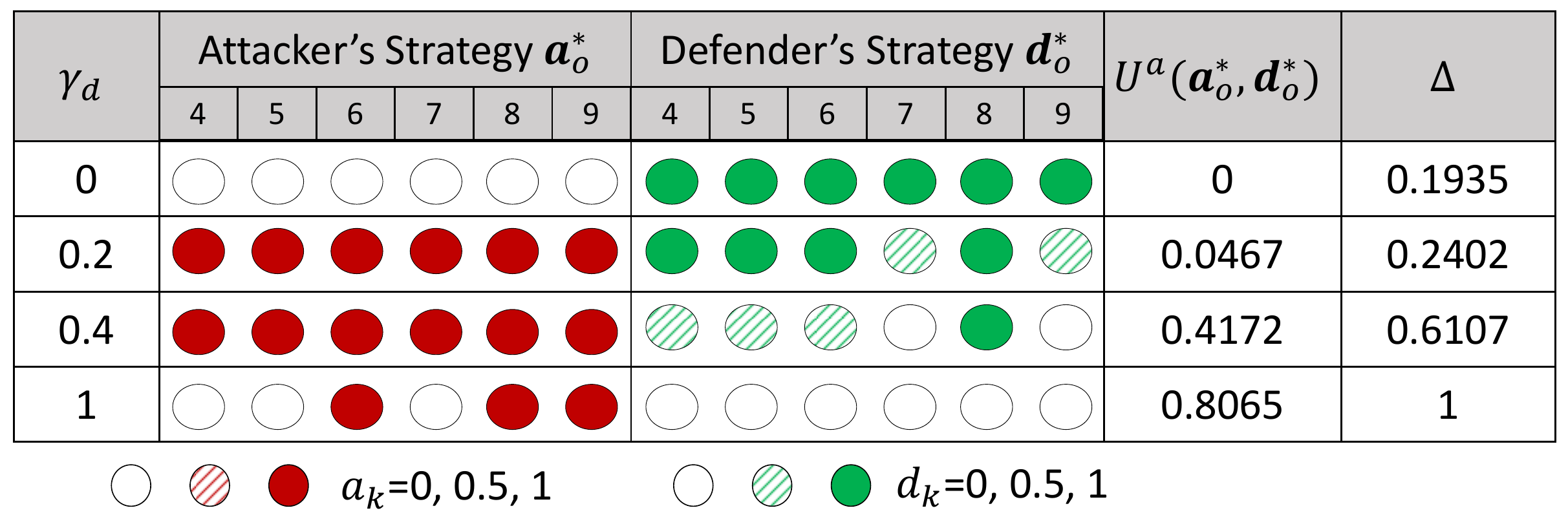}
    \caption{Player's strategies at CBSE when $\gamma_a\leq 0.15$, $L_a=L_d=3$}
    \label{fig:ga0_9}
    \vspace{-0.1in}
\end{figure}

\begin{figure}[H]
  \centering
    \includegraphics[width=0.47\textwidth]{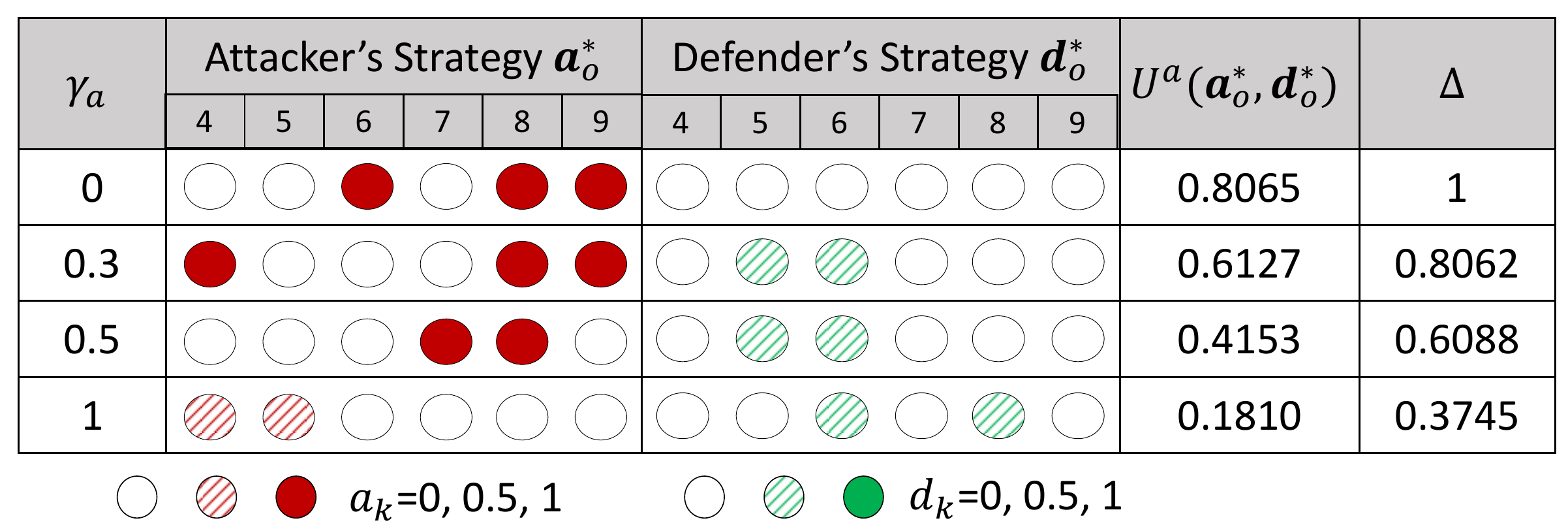}
    \caption{Player's strategies at CBSE when $\gamma_d\geq0.75$, $L_a=L_d=3$}
    \label{fig:gd1_9}
    \vspace{-0.2in}
\end{figure}

The IEEE 9-bus system has 6 load buses, which
are potential targets for the players in the proposed
SG. The nominal voltage instability index for this system is computed as $\Delta^0 = 0.1935$. In the simulation, $q_a^{k,\max}$ is determined by the covertness constraint (\ref{eq:a_cc}), and we set $q_d^{k,\max}=2$ pu, $\forall k$. It was verified that these compensations do not violate the $[0.9, 1.1]$ pu voltage range for any bus.
First, we examine the \textit{dependency of the proposed CBSG on the players' costs}. Fig. \ref{fig:Ua9_La3Ld3} shows the attacker's utility $U^a({\mathbf{a}_o^*},{\mathbf{d}_o^*})$ (\ref{eq:Ua}) at CBSE while Fig. \ref{fig:ga0_9} and \ref{fig:gd1_9} illustrate the players' strategies for varying scaled costs of attack $\gamma_a$ and protection $\gamma_d$ assuming three investment levels for each player. We observe the performance trends described in Theorem 1 (a)$\sim$(d). In Fig. \ref{fig:Ua9_La3Ld3}, the largest attacker's utility is $1-\Delta^0=0.8065$ (voltage collapse), which occurs when $\gamma_a\leq0.15$ and $\gamma_d\geq0.75$. In this case, the defender's cost per load greatly exceeds that of the attacker's, so the attacker is able to increase its reactive power demand to achieve $\Delta=1$ while the defender cannot compensate due to its limited resources. On the other hand, when the defender's cost is small ($\gamma_d\leq 0.15$), implying the defender has sufficient resources to compensate for any level of attack, the resulting $U^d({\mathbf{a}_o^*},{\mathbf{d}_o^*})=-U^a({\mathbf{a}_o^*},{\mathbf{d}_o^*})=0$ or $\Delta = \Delta^0$. Finally, we found that as $\gamma_d \rightarrow \infty$ (not shown), the defender becomes inactive. In this case, voltage collapse happens if $\gamma_a\leq 0.3$ while $\Delta=\Delta^0$ is achieved only if the attacker is also inactive ($\gamma_a>2$). By comparing these results with Fig. \ref{fig:Ua9_La3Ld3}, we conclude that strategic protection is necessary for maintaining a reliable instability index $\Delta$.

\begin{table}[h!]
\caption{``Importance" ranking of loads for the attacker and defender for the IEEE 9-bus system}
\centering
\begin{tabular}{|>{\centering\arraybackslash}m{9mm}||>{\centering\arraybackslash}m{14mm}|>{\centering\arraybackslash}m{12mm}||>{\centering\arraybackslash}m{14mm}|>{\centering\arraybackslash}m{12mm}|}
\hline
\multirow{2}{*}{Load \#} & \multicolumn{2}{c||}{Attacker} & \multicolumn{2}{c|}{Defender} \\ \cline{2-5} 
                         & $\Delta-\Delta^0$ & Ranking &  $\Delta^0-\Delta$ & Ranking \\ \hline
4    & 0.2947        & 4  &  0.0892  & 4          \\ \hline
5    & 0.2825        & 6  &  0.2379  & 1          \\ \hline
6    & 0.3040        & 1  &  0.2101  & 2          \\ \hline
7    & 0.2871        & 5  &  0.0584  & 5          \\ \hline
8    & 0.2987        & 3  &  0.1364  & 3          \\ \hline
9    & 0.3025        & 2  &  0.0257  & 6          \\ \hline
\end{tabular}
\label{tab:importance_9a}
\vspace{-0.1in}
\end{table}
Next, to illustrate the players' strategy choices, we list the \textit{``importance" ranking of loads} for both players in Table \ref{tab:importance_9a}. First, we show the increment of the instability index $\Delta-\Delta^0$ (assuming the initial value $\Delta^0$) when the reactive power demands of individual loads are increased by the maximum allowed covertness limit $q_a^{k,\max}$. The greater the increment for an individual load, the more ``important" that load is to the attacker. In addition, we illustrate the ``importance" order of the loads from the defender's perspective by examining the decrement $\Delta^0-\Delta$ when the initial value is $\Delta^0$ and the defender compensates a fixed $q_d^{k,\max}=1$ pu, $\forall k$, on a single load. Similarly, the greater the decrement, the more ``important" that load is to the defender. While Tables \ref{tab:importance_9a} shows the ``importance" ranking of the loads before the attack, i.e. when the initial $\Delta=\Delta^0$, we found that the ``importance" ranking does not depend on the initial value of $\Delta$.


In general, \textit{multiple SEs} are possible for any choice of game settings. In this example, multiple SEs occur in two regions in the range of costs shown in Fig. \ref{fig:Ua9_La3Ld3}. First, for $\gamma_d \leq 0.15$, the defender is able to invest into all loads, resulting in $\Delta=\Delta^0$, but best responses of the attacker vary, creating multiple SEs. The CBSE occurs when the attacker chooses not to act to save its cost as shown in the first row of Fig. \ref{fig:ga0_9}. Second, in the region $\gamma_d\geq 0.75$, $\gamma_a \leq 0.15$, the attacker is very strong while the defender is severely resources-limited. Thus, voltage collapse cannot be avoided. While multiple SEs exist, e.g. the attacker invests fully into all loads and the defender invests into some ``important" loads, the CBSE corresponds to the cases illustrated in the bottom row of Fig. \ref{fig:ga0_9} and top row of Fig. \ref{fig:gd1_9}. By choosing the CBSE, the defender saves cost by not acting since it cannot avoid voltage collapse while the attacker invests into its three ``important" loads, sufficient to achieve voltage collapse.
\vspace{-0.2in}
Next, we examine \textit{the effect of cost constraints on the players' investment strategies}. In Fig. \ref{fig:ga0_9} the attacker has plentiful resources ($\gamma_a\leq 0.15$). As the cost of defense per load $\gamma_d$ increases, the defender protects fewer loads and/or reduces the level of protection, thus reducing the utility $U^d({\mathbf{a}_o^*},{\mathbf{d}_o^*})$ at CBSE, or increasing $\Delta$. When $\gamma_d=0.2$, the defender targets its ``important" loads (Table \ref{tab:importance_9a}), i.e. 4, 5, 6, and 8 are fully protected ($d_k=1$) while less ``important" loads 7 and 9 are protected at half-strength ($d_k=0.5$). When $\gamma_d=0.4$, the defender's budget tightens further, and only the most ``important" four loads are protected although only the third ranked load is protected fully, revealing limitations of the load-ranking method in Table \ref{tab:importance_9a}. The latter ranking is based on attacking or defending a single load and thus is imprecise for multiple-load attack or protection scenarios due to nonlinearity of (\ref{eq:delta}).

In Fig. \ref{fig:gd1_9}, we illustrate the players' strategies at CBSE when the defender is resource-limited. For $\gamma_a\leq 0.15$, the defender is able to reduce $\Delta$ by protecting the ``important" loads 5, 6 and/or 8 at the level $d_k=0.5$. These strategies correspond to the defender's best effort under limited resources. The attacker also chooses to attack its ``important" loads, but tries to avoid investing into the loads protected by the defender. These choices are caused by the proposed game hierarchy and the nonlinear nature of the payoff function.

\begin{figure}[H]
\vspace{-0.1in}
  \centering
    \includegraphics[width=0.3\textwidth]{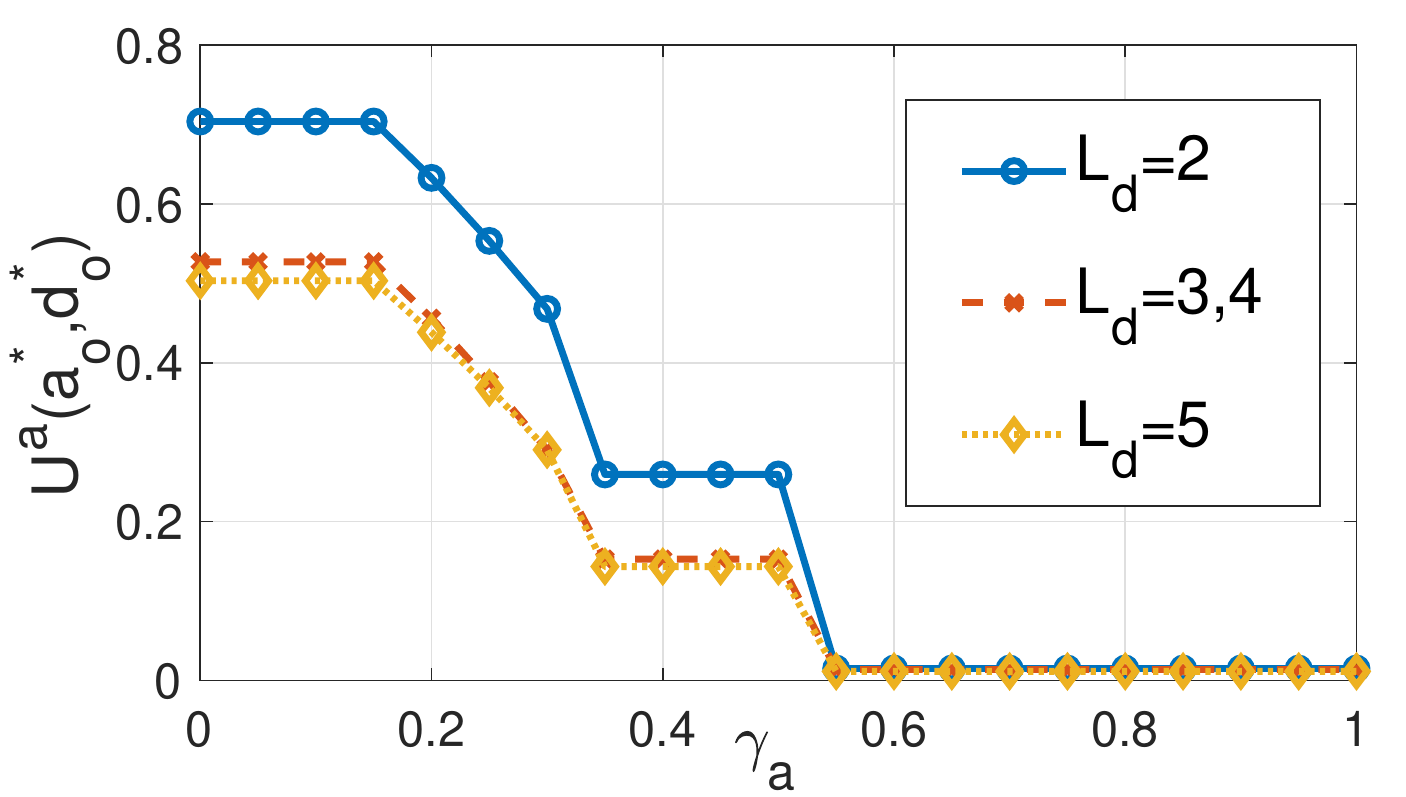}
    \caption{Attacker's utility at CBSE vs. $\gamma_a$ when $L_a=2$ and $L_d$ varies, $\gamma_d=0.5$}
    \label{fig:La2}
    \vspace{-0.1in}
\end{figure}

Next, we examine \textit{the dependency of the players' payoffs on the levels of investment} $L_a$, $L_d$. In Fig. \ref{fig:La2}, we illustrate the attacker's utility at CBSE as the defender's number of investment levels $L_d$ varies while fixing $L_a=2$. Similar simulations were performed for other scenarios, where the number of levels of one player is fixed while the other player's number of investment levels varies, and the results confirm the conclusion in Theorem 1(e). Since the game complexity scales as $L_a^K \times L_d^K$ and becomes very large even for the 9-bus system ($K=6$) as $L_a$ or $L_d$ grows, more thorough analysis of this dependency will be addressed in future work.
Finally, we compared the game described in this Section with the Individual Optimization (IO) method, where the players do not take into account the opponent’s actions or the game hierarchy \cite{AnCDCsupp,An2020thesis}. Significant losses in cost and up to 18\% loss in payoff were observed for some cost pairs for each player when using the IO method, thus underscoring the importance of strategic investment.

\subsection{CBSG for the IEEE 39-bus System}
One reason for performing a detailed analysis of our game on a relatively small power system model, such as the IEEE 9-bus system, was to demonstrate that the attacker's and defender's investment choices are mostly limited to the top few ``important" loads (from Table \ref{tab:importance_9a} and Fig. \ref{fig:ga0_9}-\ref{fig:gd1_9}). Moreover, for the IEEE 9 bus system, we compared the game above, where all loads were used (Fig. 1-4), with the SG where each player targets only its top four ``important" loads in Table I. We found that when $\gamma_a>0.5$ and $\gamma_d>0.5$, the SEs and, thus, the playoffs of the two games are exactly the same. Outside this cost range, the difference between the payoffs of the two games is at most 0.18. This result can be explained by observing that in the region $\gamma_a>0.5$ and $\gamma_d>0.5$, the attacker (or defender) has sufficient resources for attacking (or protecting) only 4 loads fully or partially with 3-level investment, and, thus, concentrates on the top four `important" loads even if other loads are included in its action set. Taking a hint from this observation, we can apply the proposed game to any larger-scale power system model over a subset of loads that includes the most ``important" loads. We expect the resulting performance to approximate closely that of the full-scale game (over all system loads) except for the cost region where one of the players is not resource-constrained. This approach reduces computational complexity significantly in practical resource-limited scenarios.

We next validate our game using important loads of the 39-bus model. This model has 29 loads, and the nominal voltage stability index is computed as $\Delta^0=0.5560$.
We first determine the most ``important" loads of the IEEE 39-bus system using the approach described in Sec.IV.A (Table \ref{tab:importance_9a}). We found that the five most ``important" loads for the attacker are $11>6>5>10>13$ while the five most ``important" loads to the defender are $7>8>5>6>11$, where $A>B$ indicates that load $A$ is ranked higher than load $B$.
\vspace{-0.1in}
\begin{figure}[H]
  \centering
    \includegraphics[width=0.35\textwidth]{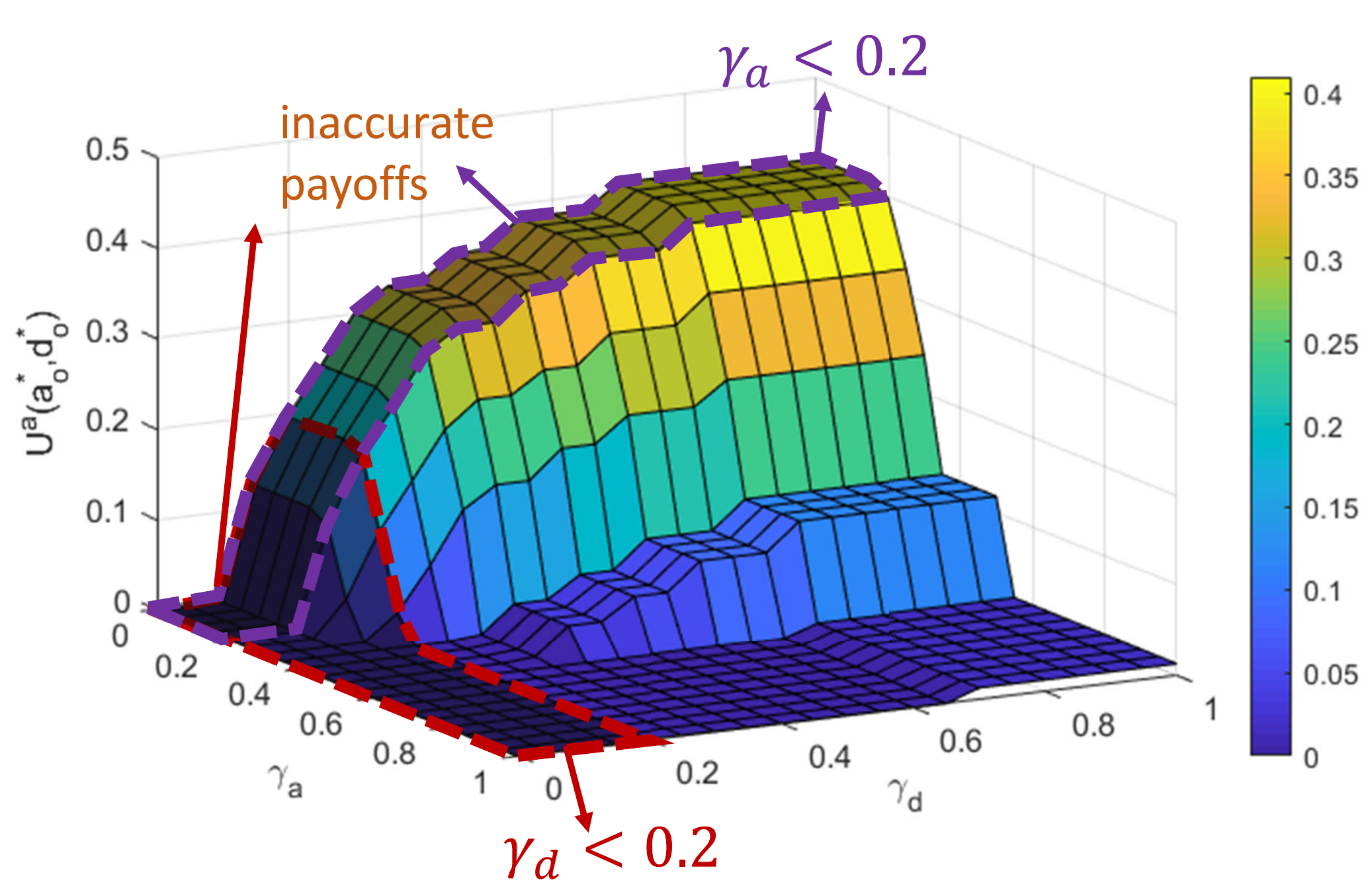}
    \caption{Attacker's utility at CBSE vs. $\gamma_a$ and $\gamma_d$ for the IEEE 39-bus system, $L_a=L_d=2$}
    \label{fig:Ua39_La2Ld2}
    \vspace{-0.1in}
\end{figure}
Based on the above analysis, we construct a CBSG over the selected subset of loads in the IEEE 39-bus system: $\{5,6,7,8,10,11,13\}$, which includes both players' five most important loads. Moreover, we assume both players employ $L_a=L_d=2$ in this pilot study since a higher number of levels increases complexity significantly. We note that in the region $\gamma_a>0.2$ and $\gamma_d>0.2$, only 5 loads can be attacked or protected with 2-level investment. Thus, in the latter region, each player concentrates most of the time on the top five ``important" loads, and the simpler SG where only the top five ``important" loads of each player are employed in the action sets is expected to closely approximate the complex game where all 29 loads are targeted. Fig. \ref{fig:Ua39_La2Ld2} shows the attacker's utility at CBSE for varying scaled costs $\gamma_a$ and $\gamma_d$. We observe the same performance trends as in Fig. \ref{fig:Ua9_La3Ld3} and Theorem 1. Note that only the region $\gamma_a>0.2$ and $\gamma_d>0.2$ provides accurate estimation of the full-scale game. Although the payoffs are inaccurate in the shaded region of Fig. \ref{fig:Ua39_La2Ld2} ($\gamma_a<0.2$ or $\gamma_d<0.2$), we are certain that voltage collapse ($\Delta=1$) occurs in this region according to Theorem 1(d), but the exact location of voltage collapse would require a full-scale game. Note that small values $\gamma_a<0.2$ or $\gamma_d<0.2$ indicate very large resources of one player, which is unlikely in practice. Finally, we observe that in the IEEE 39-bus system, the attacker is successful in raising the instability index $\Delta$ over a larger range of the cost region than in Fig. \ref{fig:Ua9_La3Ld3} since the nominal instability index of the 39-bus model is $0.5560$, which is much higher than that for the IEEE 9-bus system ($0.1935$). In other words, the 39-bus model is more ``stressed" than the 9-bus model. Nevertheless,voltage collapse is expected to occur only when the attacker has very small cost $\gamma_a$ and the defender's cost $\gamma_d$ is large. We conclude that for both examples, voltage collapse can be successfully prevented unless the defender's security resources are disproportionately limited relative to the attacker's budget.
\vspace{-0.1in}
\section{Conclusion}
We proposed a cost-based Security Investment Stackelberg game for voltage stability of a power system. In the proposed game, investment resources are allocated strategically to optimize the players' performance objectives and to save costs. It is demonstrated that voltage stability is maintained unless the defender's security budget is much lower than the attacker's budget. Future work will focus on extending the proposed game to power system models with uncertainties, scenarios where a player has limited knowledge of the opponent's resources, as well as to numerical approaches that scale well with the size of the system model.

\section*{Acknowledgment}
The authors would like to thank Pratishtha Shukla for helpful discussion on the game formulation and analysis.

\bibliographystyle{ieeetran}
\bibliography{ref_all, ref_SG, ref_power}



\section*{Supplementary material}

# Supplementary Material for: "A Stackelberg Security Investment Game for Voltage Stability of Power Systems"

Lu An, Aranya Chakrabortty, and Alexandra Duel-Hallen

## I. PROOF OF THEOREM 1

**Theorem 1(a).** A Stackelberg Equilibrium (SE) exists in a finite two-player Stackelberg game (SG) [S1].

*Proof.* Let $\Gamma^d$ denote the set of strategies $\mathbf{d}$ for the defender (leader) and $\Gamma^a$ the set of strategies $\mathbf{a}$ for the attacker (follower). Since $\Gamma^d$ and $\Gamma^a$ are finite, the game is finite, that is, the best response function $g(\mathbf{d})$ for any strategy $\mathbf{d}$ maps to a finite non-empty set (i.e. a maximum always exists for optimization (15) in the paper when $\Gamma^a$ is finite). The leader's equilibrium strategy $\mathbf{d}^*$ exists since a maximum always exists for optimization (17) when $\Gamma^d$ is finite. Similarly, the follower's equilibrium strategy, given by $\mathbf{a}^* = g(\mathbf{d}^*)$, also exists. Thus, a strategy pair $(\mathbf{a}^*, \mathbf{d}^*)$ satisfying (15), (17) exists for a finite two-player Stackelberg game. □

**Theorem 1(b).** All SEs of a zero-sum SG have the same payoffs.

*Proof.* Since the game is zero-sum, $U^a(\mathbf{a}, \mathbf{d}) = -U^d(\mathbf{a}, \mathbf{d})$. The optimization problems (15) and (17) can be combined as a mini-max problem:

$$\mathbf{d}^* = \arg\max_{\mathbf{d}} U^d(g(\mathbf{d}), \mathbf{d}) \tag{S1}$$
$$= -\arg\max_{\mathbf{d}} U^a(g(\mathbf{d}), \mathbf{d})$$
$$= \arg\min_{\mathbf{d}} U^a(g(\mathbf{d}), \mathbf{d})$$
$$= \arg\min_{\mathbf{d}} \max_{\mathbf{a}} U^a(\mathbf{a}, \mathbf{d}).$$

Denote the optimum of $\min_{\mathbf{d}} \max_{\mathbf{a}} U^a(\mathbf{a}, \mathbf{d})$ as $U_o^a$. Assume two different equilibrium strategy pairs (SEs) $(\mathbf{a}_i^*, \mathbf{d}_i^*)$ and $(\mathbf{a}_j^*, \mathbf{d}_j^*)$, $i \neq j$, are solutions of (S1). Both such strategy pairs must result in the same optimal payoffs, i.e. $U^a(\mathbf{a}_i^*, \mathbf{d}_i^*) = U^a(\mathbf{a}_j^*, \mathbf{d}_j^*) = U_o^a$, where $U_o^a$ is the solution to (S1). A similar argument can be made for the case when the game has $N$ SEs, where $N > 2$. □

**Theorem 1(c).** In a CBSG, given $L_a$ and $L_d$, the utility of each player is non-increasing with its cost per load when the opponent's cost per load is fixed.

*Proof.* Assume $L_a$ and $L_d$ are fixed. When the attacker' cost per load is $\gamma_a$ and the defender's cost per load is $\gamma_d$, the defender's utility at an SE is $U^d$. Let $\Gamma^a$ and $\Gamma^d$ denote the sets of strategies $\mathbf{a}$ and $\mathbf{d}$, respectively. Similarly, When the attacker' cost per load is $\gamma_a$, and the defender's cost per load increases to $\gamma_d'$, where $\gamma_d' > \gamma_d$, the

defender's utility at an SE is $U^{d'}$. In this case, the attacker's action space remains $\Gamma^a$ while the defender's action space for $\gamma'_d$ is denoted $\Gamma^{d'}$. From (10), it is easy to show that $\Gamma^{d'} \subset \Gamma^d$ since $\gamma'_d > \gamma_d$.

Since the game is zero-sum, $U^a(\mathbf{a},\mathbf{d}) = -U^d(\mathbf{a},\mathbf{d})$. Eq (15) can be rewritten as:

$$g(\mathbf{d}) = \arg\max_{\mathbf{a}} U^a(\mathbf{a},\mathbf{d}) \qquad (S2)$$
$$= \arg\max_{\mathbf{a}} -U^d(\mathbf{a},\mathbf{d})$$
$$= \arg\min_{\mathbf{a}} U^d(\mathbf{a},\mathbf{d}).$$

Combining (S2) with (17), the optimization of the defender becomes:

$$\mathbf{d}^* = \arg\max_{\mathbf{d}} U^d(g(\mathbf{d}),\mathbf{d}) \qquad (S3)$$
$$= \arg\max_{\mathbf{d}} \min_{\mathbf{a}} U^d(\mathbf{a},\mathbf{d}).$$

From (S3), given $\gamma_a$, the defender's utility at an SE for $\gamma_d$ is computed as $U^d = \max_{\mathbf{d}\in\Gamma^d} \min_{\mathbf{a}\in\Gamma^a} U^d(\mathbf{a},\mathbf{d})$ while its utility at an SE for $\gamma'_d$ is $U^{d'} = \max_{\mathbf{d}'\in\Gamma^{d'}} \min_{\mathbf{a}\in\Gamma^a} U^d(\mathbf{a},\mathbf{d})$. Since $\Gamma^{d'} \subset \Gamma^d$, $U^{d'} \leq U^d$. Thus, the defender's utility is non-increasing with its cost per load $\gamma_d$ when $\gamma_a$ is fixed.

A similar argument shows that when the defender's cost per load $\gamma_d$ is fixed and the attacker's cost per load $\gamma_a$ increases, the attacker's utility at an SE is non-increasing. □

**Theorem 1(d).** Given $L_a$ and $L_d$, there exist $\epsilon > 0$ and $\theta > 0$ such that when $\gamma_a < \epsilon$ while $\gamma_d > \theta$, the attacker's utility at CBSE $U^a(\mathbf{a}^*_o,\mathbf{d}^*_o) = 1 - \Delta^0$ (i.e. voltage collapse occurs). Moreover, there exists an $\alpha > 0$ such that when $\gamma_d < \alpha$, the attacker's utility at CBSE $U^a(\mathbf{a}^*_o,\mathbf{d}^*_o) = 0$ (i.e. $\Delta = \Delta^0$).

*Proof.* From (11) and (12) in the paper, the range of the voltage instability index values in this game is $[\Delta^0, 1]$ in the game. Thus, the range of the attacker's utility at an SE is $U^a(\mathbf{a}^*_o,\mathbf{d}^*_o) \in [0, 1-\Delta^0]$. First, assume that $\epsilon < 1/K$, where $K$ is the total number of loads, and $\theta > (L_d - 1)$. Then from (8) and (10), when $\gamma_a < \epsilon$ and $\gamma_d > \theta$, the defender cannot protect any loads while the attacker is able to attack all loads fully. According to the assumption (i) at the end of Section III.B, the attacker is able to cause voltage collapse when this scenario occurs. While we assumed above that $\epsilon$ and $\theta$ are sufficiently small and large, respectively, to disable the defender and fully enable the attacker, voltage collapse, i.e. $U^a(\mathbf{a}^*_o,\mathbf{d}^*_o) = 1 - \Delta^0$, can occur in other, less restrictive, scenarios as illustrated in Fig. 1.

Similarly, assume that $\alpha < 1/K$ and $\gamma_d < \alpha$. In this case, the defender can protect each load fully according to (10). Using the assumption (ii) stated at the end of Section III.B, the defender can compensate for any attack in this case, thus maintaining the nominal instability index $\Delta^0$, resulting in the attacker's utility $U^a(\mathbf{a}^*_o,\mathbf{d}^*_o) = 0$ at an SE. □

**Theorem 1(e).** When $L_d$ (or $L_a$) is increased to a number of investment levels $L'_d$ (or $L'_a$) that satisfies $L'_d - 1 = n(L_d - 1)$ (or $L'_a - 1 = n(L_a - 1)$), where $n$ is a positive integer, the defender's (or attacker's) utility does not decrease if the costs and the opponent's number of investment levels $L_a$ (or $L_d$) are fixed.

*Proof.* Assume $\gamma_a$ and $\gamma_d$ are fixed. Let $U^d$ denote the defender's utility at an SE when the attacker's number of investment levels is $L_a$ and the defender's number of investment levels is $L_d$. Similarly, when the attacker's number of investment levels is $L_a$ and the defender's number of investment levels is $L_d'$, let $U^{d'}$ denote the defender's utility at an SE.

When the defender's number of investment levels is $L_d$, the defender's actions are represented by $\mathbf{d} = [d_1, \cdots, d_k, \cdots, d_K]$, where $d_k \in \mathbf{D} = \{0, 1/(L_d-1), 2/(L_d-1), \cdots, 1\}$ denotes the defender's investment level on load $k$. Let $\Gamma^d$ denote the set of strategies $\mathbf{d}$ for the defender for this case.

When the defender's number of investment levels is increased to $L_d' - 1 = n(L_d - 1)$, where $n$ is a positive integer, the set of possible values of $d_k \in \mathbf{D}' = \{0, 1/(L_d'-1), 2/(L_d'-1), \cdots, 1\}$. Let $\Gamma^{d'}$ denote the set of strategies $\mathbf{d}'$ for the defender for this case. Since $L_d' - 1 = n(L_d - 1)$, it is easy to show that $\mathbf{D} \subset \mathbf{D}'$. Thus, $\Gamma^d \subset \Gamma^{d'}$.

When computing the SEs, the defender searches all possible actions in its action space $\Gamma^d$ (for $L_d$) or $\Gamma^{d'}$ (for $L_d'$). Since $\Gamma^d \subset \Gamma^{d'}$, $U^d \leq U^{d'}$. Thus, the defender's utility does not decrease when its number of investment levels is increased to $L_d' - 1 = n(L_d - 1)$, where $n$ is a positive integer given the costs and $L_a$ are fixed.

Similar arguments can be made for the case when the defender's number of investment levels $L_d$ is fixed and the attacker's number of investment levels $L_a$ is increased to a larger number $L_a' - 1 = n(L_a - 1)$. □

## II. COMPARISON OF THE CBSG AND INDIVIDUAL OPTIMIZATION (IO)

### A. Individual Optimization

Instead of playing the proposed SG, the attacker and the defender can choose to optimize their investment profiles individually. Moreover, when the players do not have information about their opponent, they can still optimize their utilities under given constraints. By assuming the opponent is inactive, denote the reactive power demand for the $i^\text{th}$ outcome of the attacker's investment $\mathbf{a}$ as $\mathbf{Q}_L^{a,i'} = \mathbf{Q}_L^0 + \mathbf{q}_a^i$, where $\mathbf{q}_a^i$ is defined in (6), and the defender is assumed inactive. The resulting attacker's utility for the $i^\text{th}$ outcome is expressed as:

$$U_{ind}^{a,i}(\mathbf{a}) = \left\|\mathbf{Q}_{cirt}^{-1}\mathbf{Q}_L^{a,i'}\right\|_\infty - \Delta^0, \tag{S4}$$

and, the *expected utility of the attacker* assuming the defender is inactive is given by:

$$U_{ind}^a(\mathbf{a}) = E(U_{ind}^{a,i}(\mathbf{a})) = \sum_i^{2^K} P_\mathbf{a}\left(\mathbf{O}^i\right) U_{ind}^{a,i}(\mathbf{a}), \tag{S5}$$

where $P_\mathbf{a}\left(\mathbf{O}^i\right)$ is given by (5).

The attacker's individual optimization problem is to find the investment profile $\mathbf{a}_{ind}^*$ that satisfies:

$$\mathbf{a}_{ind}^* = \arg\max_\mathbf{a} U_{ind}^a(\mathbf{a}) \tag{S6}$$

$$\text{s.t.} \quad \gamma_a \|\mathbf{a}\|_1 \leq 1, \ q_a^k \leq q_a^{k,\max}, \forall k.$$

3

Similarly, assuming the attacker is inactive, for the defender's investment $\mathbf{d}$, the resulting reactive power demand is denoted as $\mathbf{Q}_L^{d'} = \mathbf{Q}_L^0 - \mathbf{q}_d$, where $\mathbf{q}_d$ is defined in (7). The resulting defender's utility is given by:

$$U_{ind}^d(\mathbf{d}) = \Delta^0 - \left\| \mathbf{Q}_{cirt}^{-1} \mathbf{Q}_L^{d'} \right\|_\infty. \tag{S7}$$

The *defender's individual optimization* problem is to determine the individual profile $\mathbf{d}_{ind}^*$ that satisfies:

$$\mathbf{d}_{ind}^* = \arg\max_{\mathbf{d}} U_{ind}^d(\mathbf{d}) \tag{S8}$$

$$\text{s.t.} \quad \gamma_d \|\mathbf{d}\|_1 \leq 1.$$

Given $\mathbf{a}_{ind}^*$ and $\mathbf{d}_{ind}^*$ in (S6) and (S8), the actual reactive power demand of the $i^{\text{th}}$ outcome is $\mathbf{Q}_{L,ind}^{i'} = \mathbf{Q}_L^0 + \mathbf{q}_{\mathbf{a}_{ind}^*}^i - \mathbf{q}_{\mathbf{d}_{ind}^*}$, producing the instability index increment (i.e. the attacker's utility) of the individual optimization method as:

$$U_{IO}^a(\mathbf{a}_{ind}^*, \mathbf{d}_{ind}^*) = Clip\left( \sum_i^{2^K} P_{\mathbf{a}_{ind}^*}(\mathbf{O}^i) \left\| \mathbf{Q}_{cirt}^{-1} \mathbf{Q}_{L,ind}^{i'} \right\|_\infty ; (\Delta^0, 1) \right) - \Delta^0. \tag{S9}$$

Similarly, we define the defender's utility of the IO method as the $\Delta$ decrement:

$$U_{IO}^d(\mathbf{a}_{ind}^*, \mathbf{d}_{ind}^*) = -U_{IO}^a(\mathbf{a}_{ind}^*, \mathbf{d}_{ind}^*). \tag{S10}$$

## B. Numerical Results

As discussed above, in the IO method, each player chooses its optimal strategy assuming no action is taken by the opponent. On the other hand, in the SG game, each player's strategy is that player's optimal response to the opponent's action. Thus, the IO method is a suboptimal approach when the opponent is active, but it has lower complexity and does not require estimation of the opponent's capabilities. In the following, we compare the performance of the proposed CBSG game and IO.

From Fig. S1, we observe that each player's payoff can be up to 18% lower when using IO relative to playing the game for some cost pairs ($\gamma_a$, $\gamma_d$). For example, when $\gamma_a = 0.85$ and $\gamma_d = 0.65$, the defender's payoff is around 18% lower when employing the IO method than playing the game. When $\gamma_a = 0.3$ and $\gamma_d = 0.3$, the attacker loses around 16% of its payoff with IO method. Moreover, Fig. S2 compares the optimal strategies of CBSG and IO. We observe that each player targets its "important" loads in IO while in the game the attacker often prefers to avoid the loads targeted by the defender (e.g. when $\gamma_a = 0.5$ and $\gamma_d = 0.5$). Also, note that the players' costs can be significantly lower when playing the proposed CBSE than in IO since CBBI selects an SE with the lowest cost while in IO, each player invests fully up to the given budget constraints. For example, when $\gamma_a = 0.1$ and $\gamma_d = 0.1$, the attacker prefers not to act in CBSG to save cost since the defender can compensate for any action of the attacker in this case. On the other hand, in the IO method, the attacker attacks all loads, thus incurring much higher cost to obtain the same payoff as in SG in this scenario. Finally, the players' utilities (S4) and (S7) are not limited to $[\Delta^0, 1]$ since each player is not aware of the opponent's action and simply tries to modify the instability index. Thus we observe unique solutions to (S6) and (S8), resulting in a single optimal pair $(\mathbf{a}_{ind}^*, \mathbf{d}_{ind}^*)$.

In contrast, multiple SEs are possible in a SG, and the resulting CBSE has a lower cost than the corresponding IO investment strategy as Shown in Fig. S2.

By definition, an SE provides the optimal payoffs for both players. By comparing the game with IO, we evaluated the possibility of not employing the game. However, our analysis shows that each player experiences significant losses with the IO method. In particular, a player can have lower payoff and higher cost for some cost pairs in IO than in CBSG, which provides incentives for playing the game.

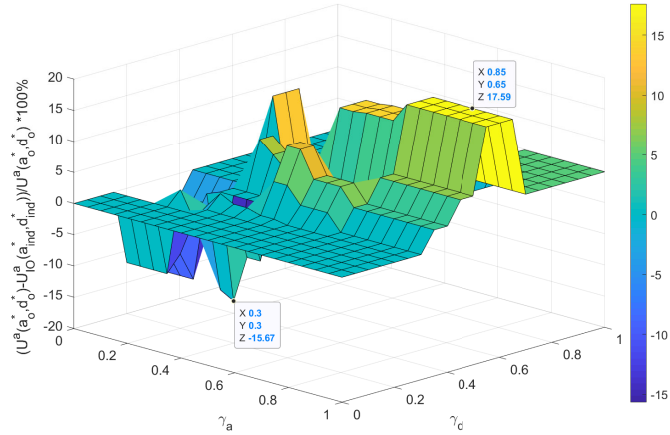

Fig. S1: Fractional difference of attacker's utility $U^a(\mathbf{a}_o^*, \mathbf{d}_o^*)$ of SG at CBSE and utility of IO $U_{IO}^a(\mathbf{a}_{ind}^*, \mathbf{d}_{ind}^*)$ (in % relative to $U^a(\mathbf{a}_o^*, \mathbf{d}_o^*)$) for the IEEE 9-bus system, $L_a = L_d = 3$

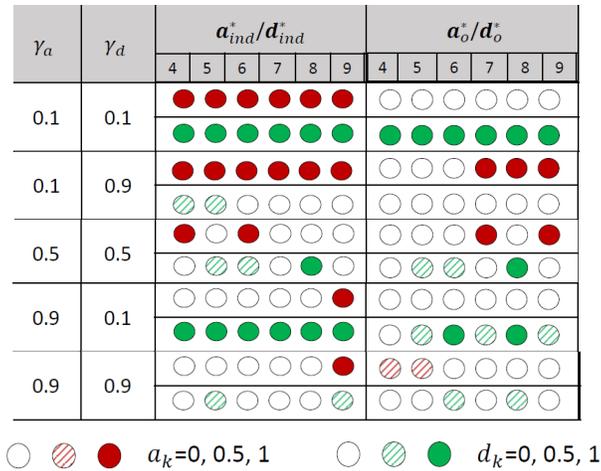

Fig. S2: Comparison of optimal strategies at CBSE and IO for the IEEE 9-bus system, $L_a = L_d = 3$


\end{document}